\documentclass[11pt,twoside]{article}
\usepackage{aussois2003,epsf,psfig}

\def\edcomment#1{\iffalse\marginpar{\raggedright\sl#1\/}\else\relax\fi}
\reversemarginpar

\newcommand{\zphot}{$z_{phot}$~}
\newcommand{\zphots}{$z_{phot}$s~}
\def\ltapprox{\raisebox{-0.5ex}{$\,\stackrel{<}{\scriptstyle\sim}\,$}}
\def\gtapprox{\raisebox{-0.5ex}{$\,\stackrel{>}{\scriptstyle\sim}\,$}}
\def\halpha{\ifmmode {\rm H{\alpha}} \else $\rm H{\alpha}$\fi}
\def\hbeta{\ifmmode {\rm H{\beta}} \else $\rm H{\beta}$\fi}
\def\oii{[O\,{\sc II}] $\lambda\lambda$3726,3728}
\def\oiii{[O\,{\sc III}] $\lambda\lambda$4959,5007}

\begin{document}
%
\title{Properties of Faint Distant Galaxies as seen 
through Gravitational Telescopes}
%
\author{Roser Pell\'o, Thierry Contini, Marie Lemoine-Busserolle,
  Johan Richard, Jean-Paul Kneib$^2$, Jean-Fran\c cois Le Borgne, 
Daniel Schaerer$^3$}
\affil{Laboratoire d'Astrophysique de l'Observatoire
  Midi-Pyr\'en\'ees, UMR5572, 14 Av. Edouard B\'elin,
F-31400 Toulouse (France)}


\author{Micol Bolzonella}
\affil{IASF-MI, via Bassini 15, I-20133 Milano, Italy}

\affil{$^2$ Astronomy Department, California Institute of Technology,
Pasadena, CA 91125} 

\affil{$^3$ Geneva Observatory, 51, Ch. des Maillettes, CH-1290
  Sauverny, Switzerland}

\label{page:first}
\begin{abstract}
This paper reviews the most recent developments related to the use of
lensing clusters of galaxies as Gravitational Telescopes in deep
Universe studies. We summarize the state of the art and the most
recent results aiming at studying the physical properties of distant 
galaxies beyond the limits of conventional spectroscopy. The
application of photometric redshift techniques in the context of
gravitational lensing is emphasized for the study of both lensing 
structures and the background population of lensed galaxies. 
A presently ongoing search for the first building blocks of galaxies
behind lensing clusters is presented and discussed. 
\end{abstract}
%
\section{Introduction}

The availability of statistically significant samples of galaxies,
from $z \sim 0$ to the largest look-back times, covering a large
region in the parameter space, is mandatory for 
constraining the cosmological scenarios of galaxy formation and
evolution. Large spectroscopic samples of galaxies at all redshifts have become
available during the last ten years, thanks to extended surveys in the
different rest-frame wavelength domains (e.g. Lilly et al. 1995; Ellis et
al. 1996; Steidel et al. 1996, 1999; Adelberger \& Steidel 2000;
Chapman et al. 2000; Pettini et al. 2001; Erb et al. 2003). 
The evolution of the overall properties of galaxies as a function of
redshift is particularly important for galaxies at z \gtapprox 1, a
redshift domain where galaxies are expected to be strongly affected 
by merging or assembly processes. 

   Clusters of galaxies acting as Gravitational Telescopes (hereafter GTs) 
constitute a particular and powerful tool in this context of high redshift surveys, 
as it was pointed out by different authors in the early nineties 
(see Fort \& Mellier 1994 for a review). 
The main advantage of such ``telescopes'' is that they 
take benefit from the large magnification factor in the core of
lensing clusters, close to the critical lines, which typically ranges
between 1 and 3 magnitudes. Thus, GTs can be used to access the most
distant and faintest population of galaxies, extending the current
surveys towards intrinsically fainter samples of galaxies,
allowing to probe the physical properties of the faint end of the 
luminosity function at moderate and high redshifts. Indeed,
GTs have been succesfully used during the last ten years to perform
detailed studies of distant galaxies at different wavelengths, from UV
to submillimeter. A particular and recent application of GTs is the
identification and study of extremely distant galaxies, including
the search for the first galaxies formed in the early
Universe, up to redshifts of the order of $z \sim 10-20$.

\begin{figure}
\centerline{\psfig{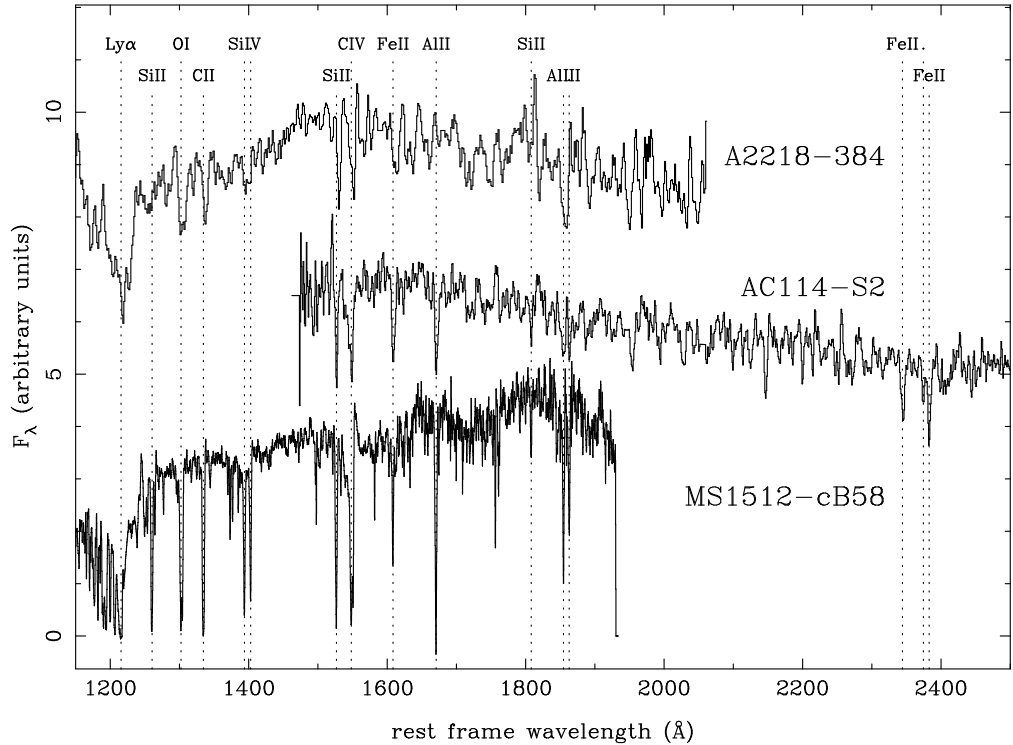}}
\caption{Spectra of three emblematic lensed LBGs
identified in the nineties, showing the most prominent absorption
lines. From top to bottom: \#384 in A2218 (z$=2.515$, Ebbels et
al. 1996), S2-AC114 (z$=1.867$, Smail et al. 1995; Campusano et
al. 2001), and cB58-MS1512+36 (z=$2.7$, Yee et al. 1996; Pettini et
al. 2000). }
\label{spectres}
\end{figure}

   The plan of the paper is as follows. Section~\ref{TGs} presents a
summary of the latest developments concerning the use of cluster
lenses as Gravitational Telescopes. We summarize the state of the art
and the recent results aiming at studying the physical
properties of distant galaxies beyond the limits of conventional
spectroscopy. In Section~\ref{photoz}, we briefly describe the
photometric redshift approach, as a tool in deep photometric studies,
and in particular in the context of gravitational lensing. We
emphasize the use of such a tool for the study of both lensing
structures and the background population of lensed galaxies.
We discuss in Section~\ref{popIII} on the use of Gravitational
Telescopes to identify the first building blocks of galaxies, also
known as Population III sources. A presently ongoing search for PopIII 
candidates behind lensing clusters is briefly presented. Conclusions are
given in Section~\ref{conclusions}.

\section{Physics of galaxies beyond the limits of
 conventional spectroscopic samples. State of the art}
\label{TGs}

\subsection{Lensing Clusters as Gravitational Telescopes}

\begin{figure}
\centerline{\psfig{file=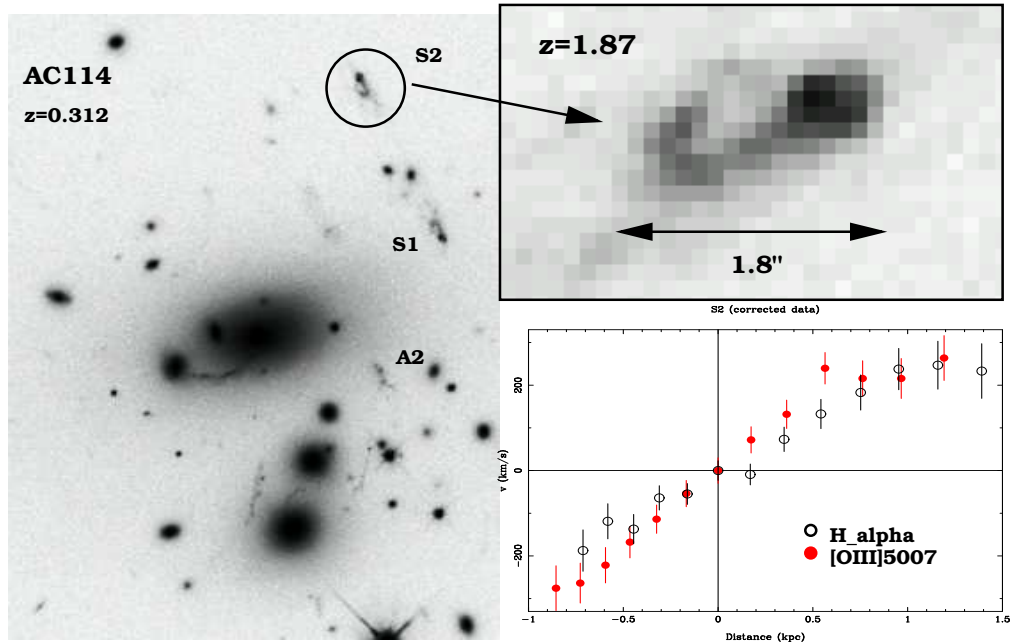,width=1.0\textwidth,angle=0}}
\caption{Velocity field in AC114-S2 (z$=1.867$, from Lemoine-Busserolle
et al. 2003). Lens corrected velocity profile (in km/s) versus distance 
measured on the 2D spectra from the central velocities of {\rm H${\alpha}$}
(open dots) and [O\,{\sc III}] $\lambda5007$ (full dots) emission
lines. A velocity gradient of $\pm 240$ km/s is measured, 
which yields a dynamical mass of $\sim 1.3 \times 10^{10}$ M$_{\odot}$ within
the inner 1 kpc radius.}
\label{ac114_rot}
\end{figure}

From the point of view of galaxy evolution studies, GTs can be
considered as an alternative way to blank fields to investigate the
properties of distant galaxies. 
GTs allow to construct and to study independent samples of 
high-z galaxies, less biased in luminosity than standard
magnitude-limited field samples, thanks to the magnification factor.
For a given limiting magnitude or flux,
lensed samples will complement the currently available field 
surveys towards the faint end of the luminosity function, and towards
higher limits in redshift. 
It is worth noting that only lensing clusters with fairly well 
constrained mass distributions can be used as efficient GTs for galaxy
evolution studies. In order to optimize the surveys, the expected 2D 
surface density distribution of arclets in clusters can be 
estimated for lenses with well known mass-distributions. Conversely, 
the lens-corrected distribution of arclets can be easily retrieved 
from the cluster cross section (e.g. Kneib et al. 1994, 
Bezecourt et al. 1998). An interesting property of
GTs is that they conserve the surface brightness and the spectral 
energy distribution (SED) of lensed galaxies. Besides the spatial 
magnification, sources (or regions of sources) smaller than the 
seeing will effectively gain in surface brightness. These two effects
contribute to increase the interest of GTs in spectroscopic studies.

The use of lensing clusters as genuine GTs is a relatively new research
domain. Historically, it started 
with the spectroscopic confirmation of the first highly amplified
source at $z=0.724$ in A370 (Soucail et al. 1988), and the
identification of the spectacular blue arc in Cl2244-02 as a high-z
lensed galaxy at $z \sim 2$ (Mellier et al. 1991). The power of GTs to
study the spectrophotometrical and morphological properties of
high-$z$ galaxies was emphasized by different authors in the 
nineties (e.g. Smail et al. 1993; Smail et al. 1996; Kneib et
al. 1996). 
\begin{figure}
\centerline{
\psfig{file=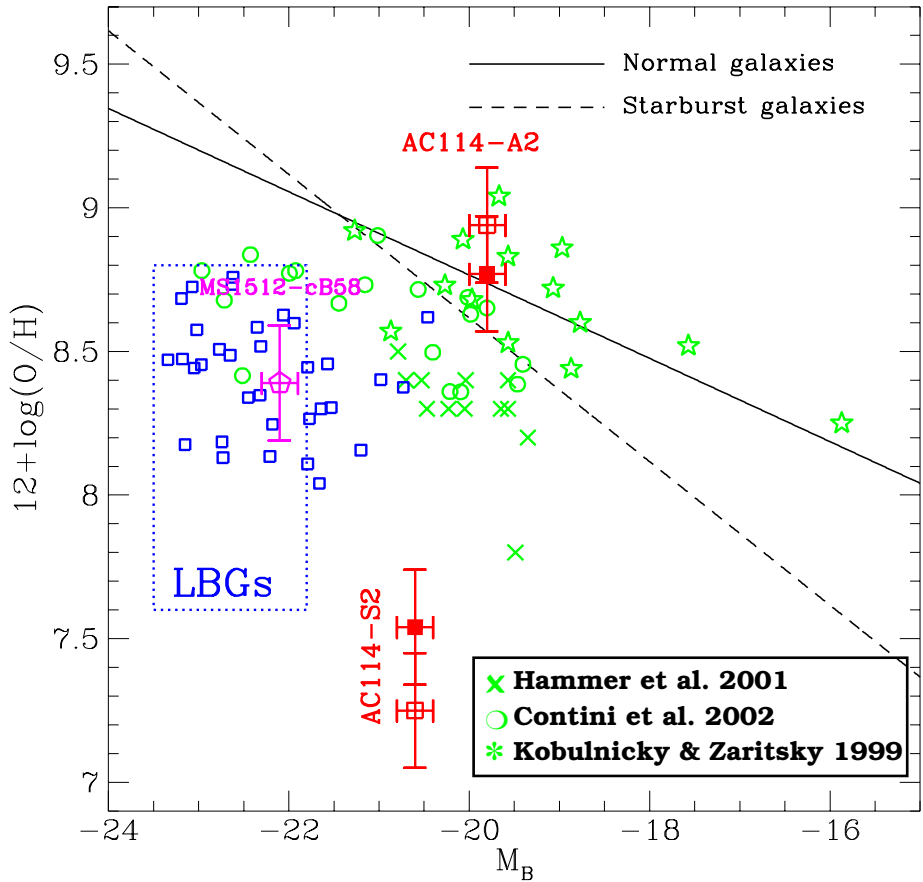,width=0.950\textwidth}}
\caption{Oxygen abundance as a function of M$_B$, showing
the location of the two $z \sim 1.9$ lensed galaxies S2 and A2 in
AC114 (from Lemoine-Busserolle et al. 2003), with (filled squares) and 
without (empty squares) extinction correction. The
metallicity--luminosity relations for nearby ``normal'' (solid line;
Kobulnicky \& Zaritsky 1999) and starburst (dashed line; Mouhcine \& Contini 2002;
Melbourne \& Salzer 2002) galaxies are also shown, together with 
different samples of intermediate-redshift galaxies. Blue squares are
high-redshift ($1.4<z<3.4$) galaxies with O/H
derived from rest-frame UV spectra by Mehlert et al. (2002).
LBGs at $z \sim 3$
are shown as a blue box encompassing the range of O/H and M$_B$
derived for these objects by Pettini et al. (2001).
}
\label{z2gal_1}
\end{figure}

One of the first detailed studies on the physical properties of
high-$z$ lensed galaxies was carried out by Ebbels et al. (1996), who 
studied the optical spectrum (obtained with the 4m telescope WHT) 
and optical-infrared colours of a star-forming galaxy at $z$=2.515 
behind A2218. This galaxy exhibits the same properties as typical 
Lyman Break Galaxies (LBGs, Steidel et al. 1996) in terms of colors,
luminosity and star formation rate (see also Leitherer 1999). 
Another emblematic source is the luminous fold arc cB58, a z=2.7
star-forming galaxy behind the EMSS cluster MS1512+36, which is probably
the best studied LBG nowadays (Yee et al. 1996; Seitz
et al. 1998; Pettini et al. 2000; Teplitz al. 2000). Four galaxies
with $3.3 \le z \le 4$ were found in the rich cluster
Cl0939+47 by Trager et al. (1997), three of them at $z \sim 4$.
Because of the weakness of the common stellar wind features in the
spectra of these objects, the authors concluded that they have
subsolar metallicities, their properties beeing quite similar to the
already known ones in LBGs at $z \sim 3$. Because these objects,
revealed by the GT, are close together both in redshift and space, 
they were identified by the authors as low-metallicity Population 
II protospheroid clumps, in other words, the progenitors of today's spheroids. 
Another close look towards a star-forming region at $z \sim 4$ was provided
by the two lensed galaxies found in the cluster Abell 2390, at
$z=4.05$ (Frye \& Broadhurst 1998; Pell\'o et al. 1999). 
These galaxies are relatively bright compared to LBGs at $z \ge 3$, or
to the  $z \sim 4$ sources in Cl0939+47. They have subsolar
metallicities according to the overall fit of their SEDs, 
and they are clumpy and elongated, as revealed by the lens 
inversion procedure. They are located $\sim 100
h^{-1}_{50}$ kpc apart from each other on the source plane, with
similar orientation. Again, these results point towards a merging
process of protospheroidal clumps, revealed by the GT. 
One of the highest redshift objects known in the nineties was also discovered
thanks to the GT approach. This object was found in the cluster
CL1358+62, with $z=4.92$ (Franx et al. 1997). Its spectral 
energy distribution is quite blue, as expected for a young galaxy with
the observed restframe UV flux, but Soifer et al. (1998) results pointed
towards an important absorption by dust in this object, the first
direct evidence for such effect at $z \sim 5$.

GTs have been also used to address the morphological properties of
high-$z$ lensed galaxies, in particular using HST
images after its refurbishment. Gravitational arcs on these images 
are detected with diffraction-limited resolution. Therefore, 
morphological criteria have 
been introduced to identify lensed sources (Kneib et al. 1996). 
Smail et al. (1996) used this approach to discuss the
properties of a large number of arcs in well known clusters, and they
concluded to a strong evolution in the size of star-forming galaxies
between $z \sim 1$ and the local universe, distant sources beeing 1.5
to 2 times smaller than local ones. 
A detailed multi-wavelength study of a ring-like starburst galaxy at
$z \sim 1$, hosting a central active nucleus, was presented by Soucail
et al. (1999) as a result of their survey in the field of A370. The
comparison between this galaxy and other similar galaxies in the local
universe was possible thanks to lens inversion. Also, as mentioned
above, several studies of the close environment of $z \sim 4$ lensed
sources were conducted with GTs.

When lensed galaxies are selected for subsequent spectroscopic
studies, they are usually chosen close to the high-$z$ critical 
lines in order to obtain the largest magnification (typically $\sim 2$
magnitudes). Number of high-$z$ lensed galaxies observed during the
nineties were discovered serendipitously, but efficient selection
procedures can be defined, based on photometric or lensing criteria.
Therefore, number of
recent surveys have started using GTs to constitute large samples of
intrisically faint objects at intermediate and high redshifts, using
the capabilities of present day multi-object spectroscopy facilities
in the optical domain. Among them,
Campusano et al. (2001) presented a spectroscopic survey of 10 faint
lensed galaxies in the core of the cluster AC114
($z_{cluster}$=0.312), with $0.7 \ltapprox z \ltapprox 3.5$. 
Of particular interest are five objects (A, B, C, S and E) 
with spectroscopic redshift
$2.0 \ltapprox z \ltapprox 3.5 $, which are between 0.5 and 1.5
magnitudes fainter than the limiting magnitude in the Steidel et al.
(1999) sample of LBGs at similar redshifts. 
In a recent paper, Mehlert et al. (2002) discussed on the evidence of
chemical evolution in the spectra of high-$z$ galaxies, by comparing a
local sample of galaxies with a VLT/FORS spectroscopic sample at $1.4 \ltapprox
z \ltapprox 3.4$, using the equivalent width of CIV as a
metallicity indicator. Part of their sample is constituted by high-$z$
lensed galaxies observed in the field of 1E0657-558 (Mehlert et
al. 2001), with very large magnifications ($\sim 3$ mags.) allowing to
compare the abundances measured in faint galaxies in the local
and high-$z$ universe. These authors conclude that the major trend in
their sample is a metallicity enhancement towards low-$z$ galaxies,
with no overall dependence of metallicity with galaxy
luminosity. These results have been derived from a small number of
intrinsically faint galaxies, and they have to be confirmed by larger
samples of lensed galaxies at high-$z$.

\subsection{A few words about sample selection criteria}

During the last ten years, samples of high-$z$ lensed galaxies have
been pre-selected in the core of lensing clusters using different
techniques. The fraction of high-$z$ galaxies in magnitude-limited
samples is relatively small, even in deep photometric surveys, but
the same photometric selection techniques usually applied to blank field
studies can be adopted to select high-$z$ galaxies behind lensing
clusters. Broadband color criteria have been successfully used to
identify galaxies at redshifts higher than $z \sim 2$ (e.g. Steidel et
al 1995; Erb et al. 2003), and more general photometric redshift
techniques are likely to produce equally usefull results in this
context (see Section~\ref{photoz}).  
In the paper by Campusano et al. (2001), lensed galaxies were
selected in areas close to the high-$z$ critical lines predicted by the
gravitational lens model of Natarajan et al (1998), using a
combination of both lensing and photometric redshift criteria. All
galaxies targetted by the spectroscopic follow-up were found to be
background galaxies with redshifts values within the $0.7 \le z \le 3.5$
interval. 

\begin{figure}
\centerline{
\psfig{file=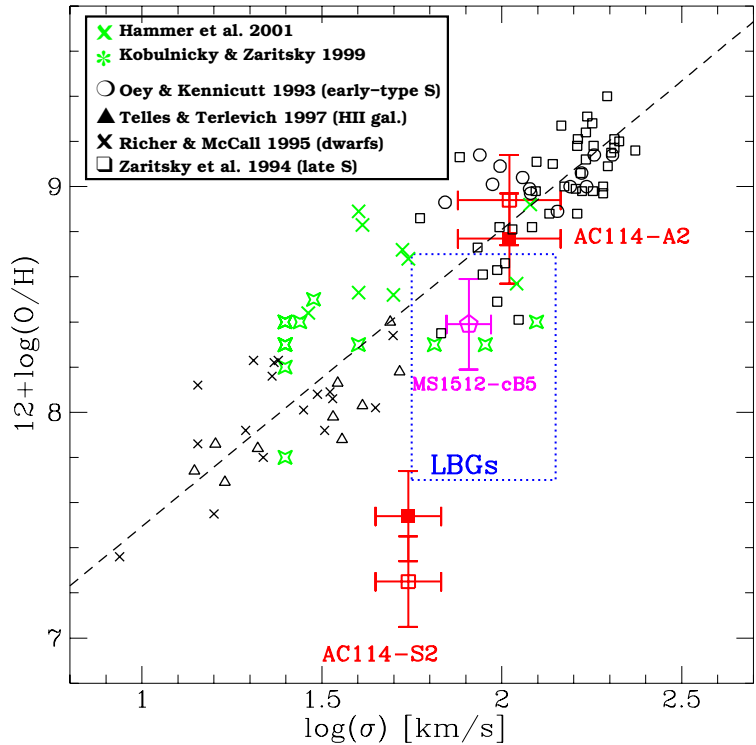,width=0.80\textwidth}}
\caption{ Same caption as in Figure~\ref{z2gal_1}: Oxygen abundance as
a function of line-of-sight velocity dispersion ($\sigma$), from
Lemoine-Busserolle et al. (2003).   
The dashed line is a linear fit to nearby and intermediate-z samples.
LBGs at $z \sim 3$
are shown as a blue box encompassing the range of O/H and $\sigma$ 
derived for these objects by Pettini et al. (2001).
}
\label{z2gal_2}
\end{figure}

Another selection criterium was proposed by Kneib et. al. (1994) and
Kneib et. al. (1996) based 
on the inversion of the lensing equations and the prediction of redshifts for
individual arclets. This is a purely geometrical criterium, based on
the averaged shape of lensed sources, and the accurate
measurement of the shape and orientation of arclets on high-resolution
HST images, although the first applications were obtained on
ground-based data (Kneib et. al. 1994). This technique produces 
statistical results on the redshift distribution of background
sources. The spectroscopic verification of this technique on
statistical basis was obtained by Ebbels et al. (1998) in the field of
A2218. Bezecourt et al. (1999) used this lens inversion method to derive
the comoving star formation rate in the range 0.5$<$z$<$2 from UV
(WFPC2/F336W) images of the cluster lens A370.

\subsection{The near-IR window}

Observations in the near-IR window have become widespread during the
last years. Indeed, without this wavelength domain, only star forming 
galaxies will be 
entering the high-z samples of galaxies defined in the visible bands. 
Deep observations at longer
wavelengths are needed to probe stellar masses, thus constraining the
cosmic star formation history. The near-IR window is barely affected
by the presence of dust extinction or starbursts, and k-corrections
are reduced compared to optical bands. This window is required to
consistently follow the stellar population contributing to the flux
beyond $\lambda \ge $4000 \AA\ from $z \sim 0$ all the way to redshift
4, and to map the star-forming and AGN activity up to the highest
redshifts ($z \sim 10$). The evolution of the overall near-IR
properties of galaxies (Luminosity Functions, redshift distribution of 
near-IR samples, color distributions, ...) as a function of redshift 
constitutes a powerful test to discriminate between the different 
scenarios of galaxy formation, in particular for galaxies at $1 \le 
z \le 2$, a redshift domain where GTs are particularly efficient. 
The Hubble Deep Field data have shown that the fraction of
galaxies at $z>1$ reaches 50\%\ typically at K$\sim$22.5-23. This is
indeed a key redshift domain where most galaxies were affected by
major merging/assembly processes. In the case of early type galaxies,
the formation of massive objects seems to took place at
redshifts $z \ge 2$, i.e. the majority of their masses were already
assembled and they only underwent minor mergers at lower redshifts.
The main spectral signatures of such assembly
process could be identified and studied through optical {\it and}
near-IR data, the later beeing crucial to determine the
photometric redshifts of galaxies within the redshift interval $ 1
\ltapprox z \ltapprox 2$ (see Section~\ref{photoz}).

The first results on a recent photometric survey for Extremely Red Objects (EROs),
undertaken with the GT on a sample of 10 clusters at $z \sim 0.2$, was
recently  published by Smith et al. (2002). 60 EROs were found with
$(R-K)\geq 5.3$ down to $K=20.6$. After correction for lensing
effects, number counts are found to flatten for magnitudes fainter
than $K\sim19$--20, a result which could be due to  
the transition from a population of EROs dominated by evolved
galaxies at $z\sim1$--2 ($K \ltapprox 19$--20) to one dominated by 
dusty starbursts at $z>1$ ($K \gtapprox 19$--20). 

   The recent development of near-IR spectrographs on 10m class
telescopes has allowed the study of the rest-frame optical properties
of high redshift galaxies. Indeed, near-IR spectroscopy allows to
access the most relevant emission lines (\halpha, \hbeta, \oii, \oiii,
...), in order to probe the physical properties of galaxies (SFR,
reddening, metallicity, kinematics, virial mass, etc), all the way
from the local universe to $z \sim 4$, using the same parameter space
and indicators. The pioneering work by Pettini and collaborators 
(1998, 2001) has shown that the
rest-frame optical properties of the brightest Lyman break galaxies (LBGs)
at $z \sim 3$ are relatively uniform. A new sample of 16 LBGs at $2
\le z \le 2.6$ has been recently published by Erb et al. (2003) and, in
this case, significant differences are found in the kinematics
of galaxies at $z \sim 2$ compared to $z \sim 3$. In a recent paper,
Lemoine-Busserolle et al. (2003) have presented the firsts results on
their near-IR spectroscopic survey of high-redshift magnified galaxies
at $z \sim 2$ using GTs. Two $z \sim 1.9$ lensed star-forming galaxies have been
studied in the field of the cluster AC 114 (S2 and A2), 1 to 2 magnitudes fainter
than LBGs at $z \sim 3$. Thanks to the large magnification factors
obtained for these objects, number of rest-frame optical
emission lines have been measured from [O\,{\sc II}]$\lambda$3727 to
\halpha+[N\,{\sc II}]$\lambda$6584. Before this paper by
Lemoine-Busserolle et al. (2003), there was only one LBG for which
chemical abundances have been determined with some degree of
confidence: the gravitationally lensed galaxy MS 1512-cB58 at
$z=2.729$ (Teplitz et al. 2000; Pettini et
al. 2002a). Figure~\ref{z2gal_1} displays the metallicity--luminosity
relationship and the metallicity versus line-of-sight velocity
dispersion ($\sigma$, which corresponds to a crude mass--metallicity
sequence) for these lensed galaxies compared to reference samples of
galaxies at different redshifts. The
results obtained up to now on this dramatically small sample suggest 
that high-$z$ objects of different luminosities could have quite 
different histories of star formation. The sample of LBGs
observed in the near-IR aiming to explore the physical properties of
galaxies is still poor. The contribution of GTs to these detailed studies is
already significant, and there is a good chance that it increases in
the following years.  

\subsection{Extending the wavelength domain}

In recent years, GT studies of lensed galaxies have successfully extended to other
wavelength windows:

\begin{itemize}

\item In the mid-infrared (MIR), ultra-deep observations through the
lensing  cluster A2390 were obtained with ISOCAM/ISO at 7 and 15 $\mu$
m (Altieri et al. 1999). A large number of sources were identified as
high-z lensed galaxies, and number counts computed to unprecedent
depth. These results ruled out non-evolutionary models and favour of
a very strong evolution (see also Metcalfe, L., this Conference). 

\item In the sub-millimeter domain, the pioneering work by Smail et
al. (1997) with SCUBA/JCMT, on the lensing clusters A370 and Cl2244-02,
showed that the majority of lensed sources detected at 450 and 850
$\mu$ m lie at redshifts above $z \sim 1$. In more recent papers (Smail et
al. 1998; Blain et al. 1998, 1999; Ivison et al. 2000; Smail et
al. 2002; see also Kneib,
this Conference), GTs have been used to trace the dust obscured 
star-formation activity at high-$z$, thanks to enlarged samples of
faint submillimeter-selected sources. 
It appears from these developments that the bulk of the 850$\mu$m background
radiation could be originated in distant ultraluminous galaxies, and
that a large fraction of $z \ltapprox 5$ sources could be missing from
optical surveys. 

\end{itemize}

\subsection{Identification and study of extremely high-$z$ sources}

Nowadays, one of the most exciting challenges for GTs is the
identification and study of extremely high-$z$ sources. Number of
devoted surveys, using different techniques, are specifically tuned up
to search for $z \gtapprox 5$ galaxies using GTs (see also Kneib, this
Conference). 

    The first results of these systematic
searches for $z \gtapprox 5$ galaxies in GTs have appeared during the last
two years. A z$=5.576$ double-imaged galaxy has been detected in the
field of A2218 (Ellis et al. 2001). This source corresponds to an
extremely faint and compact object (I$\sim 30$), highly magnified
(about a factor of $\sim 30$), with a SED dominated
by the Ly$\alpha$ emission line. According to the authors, this source
could be a $\sim 10^6 M_{\odot}$ stellar system undergoing its first
generation of stars. Another galaxy at z$=6.56$ has been identified in
the field of the cluster A370 by Hu et al. (2002). In this case, the
magnification is a factor of $\sim 4.5$ for this source, which sets
strong constraints on the reionization epoch. 
We present in Section~\ref{popIII} a particular ongoing project 
devoted to the identification of Population III candidates in lensing
clusters. 

\section{Photometric Redshifts and Gravitational Telescopes}
\label{photoz}

\subsection{Photometric Redshifts, a unique tool in deep universe
studies}
\label{photoz-intro}

Today, photometric redshifts (hereafter $z_{phot}$s) constitute 
a well established tool in deep universe studies. This technique was
originally proposed by Baum (1962), to measure the redshifts of
elliptical galaxies in distant clusters, and rediscovered later 
by several authors in the eighties (Couch et al. 1983,
Koo 1985), this time applied to relatively low-redshift samples 
of galaxies observed in the $\sim 4000$ to $8000$\,\AA\ domain. 
The interest for this technique has increased later in the nineties, 
with the development of extremely large and/or deep field surveys.

Indeed, deep photometric galaxy samples have become available during 
the last decade, in particular the observations by HST of the 
HDF-N (Williams et al. 1996) and HDF-S (Casertano et al. 2000), 
and the coordinated complementary observations from the ground
at near-IR wavelengths (Dickinson et al. 2001; Labb\'e et al. 2003).  
At the same time, reliable photometric redshift techniques have been 
developed by different authors, 
allowing distance estimates of faint galaxies for 
which no spectroscopic redshifts can
be obtained nowadays, even with the most powerful telescopes (e.g.\
Connolly et al. 1997; Wang et al. 1998; Giallongo et
al.\ 1998; Fern\'andez-Soto et al.\ 1999; Arnouts et
al.\ 1999; Bolzonella et al.\ 2000; Furusawa et al. 2000; 
Rodighiero et al.\ 2001; Rudnick et al. 2001;
Le Borgne \& Rocca-Volmerange 2002; 
and the references therein).

\subsection{Methods and accuracy}

There are basically two different techniques for measuring \zphots:
the so-called empirical training set method, and
the SED fitting method. The first approach was originally proposed 
by Connolly et al. (1995, 1997). It consists on deriving empirically the 
relationship between magnitudes and redshifts using a subsample of objects 
with measured spectroscopic redshifts, i.e. the training set.
A slightly modified version of this method was proposed by Wang et
al. (1998) to derive redshifts in the HDF-N by means of a linear
function of colours. The main advantages of this method are the small
dispersion attained, even with a small number of filters, and
the fact that it does not make use of any assumption concerning galaxy
spectra or galaxy evolution, thus bypassing the problem of a poor 
knowledge of high redshift spectra.
The main disadvantage is that the empirical relation between
magnitudes and redshifts has to be recomputed for each filter set and
survey, on a well suited spectroscopic subsample, thus making this
approach not very flexible. In addition,
the redshift range between $1.4$ and $2.2$ had been hardly
reached by spectroscopy up to now, because of the lack of strong
spectral features accessible to optical spectrographs. And 
efficient multi-object near-IR spectrographs are still missing on 10m
class telescopes. Thus, no reliable empirical relation can be obtained 
in this interval.

In the SED fitting procedure, the photometric redshift of a 
given object simply corresponds to the best fit
of its photometric SED by the set of template spectra, obtained
through a standard minimization procedure. This method 
obviously relies on the fit of the overall shape of spectra and
the detection of strong spectral features, such as the $4000$\,\AA\ break,
the Lyman decrement or particularly strong emission lines.
In the nineties, this method has been applied to the HDF data, using
either observed or synthetic template spectra (e.g. Mobasher et al. 1996;
Lanzetta et al. 1996; Gwyn \& Hartwick 1996; Sawicki et al. 1997;
Giallongo et al. 1998; Fern\'andez-Soto et al. 1999; Arnouts et
al. 1999; Furusawa et al. 2000; Bolzonella et al. 2000; Rudnick et al. 2001). 
A combination of this method with the Bayesian marginalization
introducing a prior probability was successfully proposed by
Ben\'{\i}tez (2000) (see also Coe, this Conference).
The advantage of the SED fitting procedure is that no (large) spectroscopic
training set is needed, although a control set is highly recommended.

\begin{figure}
\centerline{\psfig{file=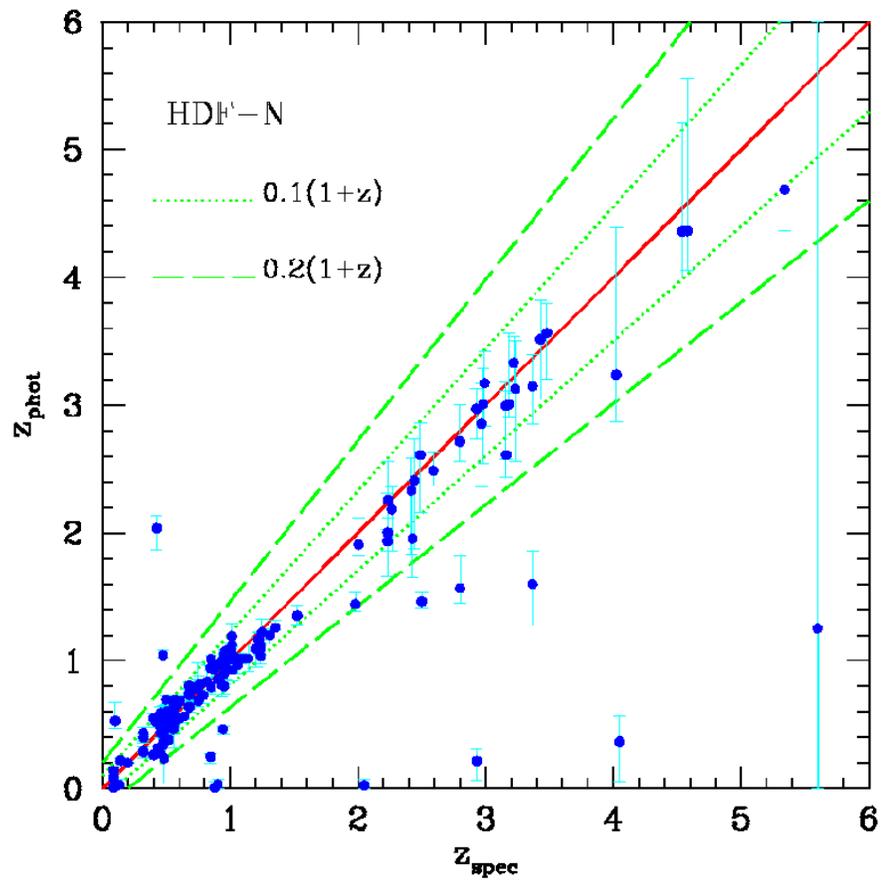,width=0.90\textwidth,angle=0}}
\caption{Comparison between spectroscopic and photometric redshifts in
the HDF fields (from Bolzonella et al. 2000). Photometric redshifts
and  $3\sigma$ error bars were computed with {\it hyperz\/}. Most of
the catastrophic identifications correspond to degenerate solutions.}
\label{HDF}
\end{figure}

In order to estimate the accuracy of \zphots, a crucial test is
the comparison between the photometric and the spectroscopic redshifts
obtained on a restricted subsample of relatively bright objects, even
if there is no guarantee that we are dealing with the same type of 
objects in the photometric (faint) and the spectroscopic (bright) samples.
A complete discussion on the expected accuracy of \zphots, based on
simulations and available spectroscopy in the HDFs, was presented by
Bolzonella et al. (2000). In this paper the authors used the public code
{\it hyperz\/}, which adopts a standard SED fitting method, but most
results should be generalized to other $z_{phot}$ tools.
{\it hyperz\/} is available on the web at {\tt  
http://webast.ast.obs-mip.fr/hyperz\/}. 
Figure~\ref{HDF} displays a comparison between spectroscopic and 
photometric redshifts in the HDF fields.
The rms uncertainties on z$_{phot}$ are typically $\sigma_z
\sim 0.05 (1+z)$, and up to $\sigma_z
\sim 0.1 (1+z)$, depending on the photometric depth, the redshift of
the sources and the spectral coverage of the filter set. 
Without near-IR data, errors on z$_{phot}$ are
relatively higher in the redshift range $ 1 \ltapprox z \ltapprox 2$,
due to the lack of strong signatures in the observed SED between
3500\AA\ and 9000\AA. The 4000\AA\ break span the near-IR domain at
redshifts between 1.2 and 5. The gain using near-IR photometry 
is also sensible for the determination of rough spectral types, such as
early versus late type galaxies, or star-galaxy-qso discrimination. A
complete discussion can be found in Bolzonella et al. (2000).

\begin{figure*}
\centerline{\psfig{file=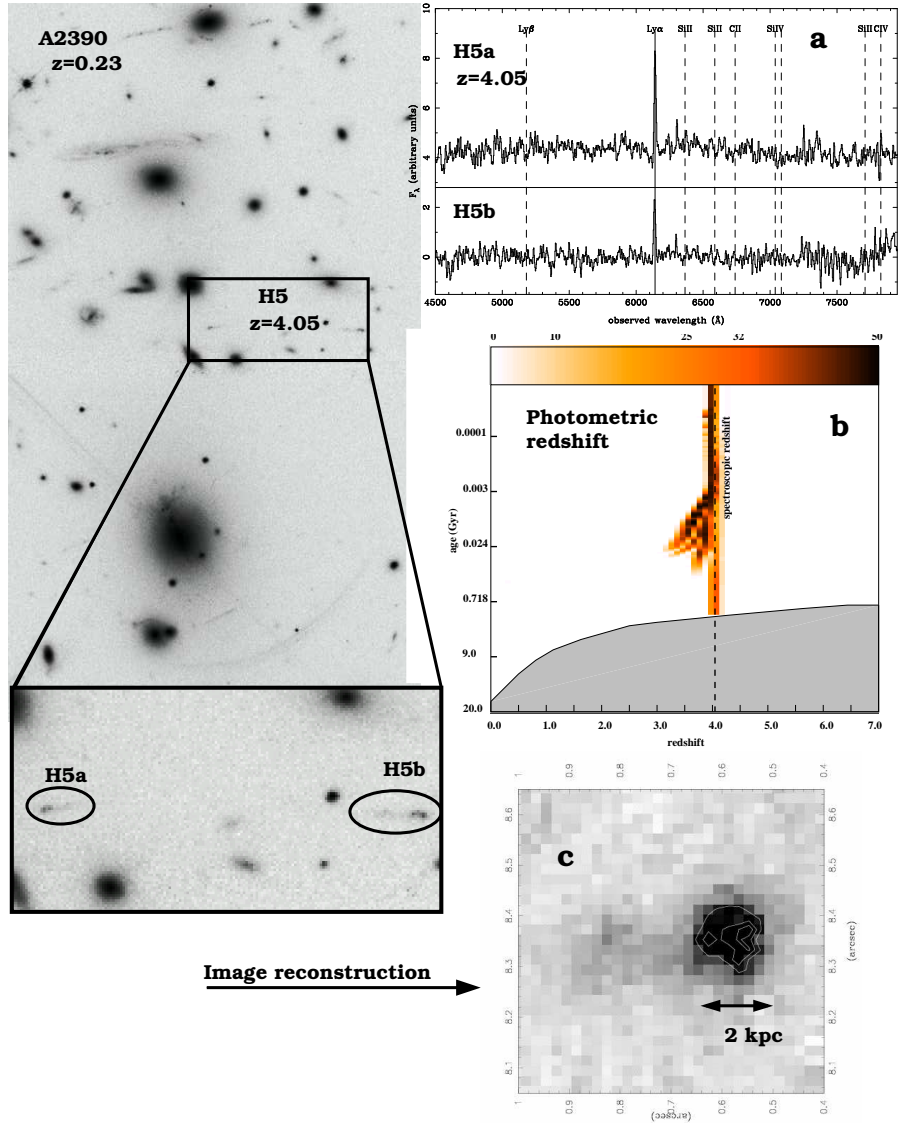,width=0.90\textwidth,angle=0}}
\caption{HST image of the cluster Abell 2390 in I band (F812W),
showing the identification of the double-imaged source H5 at z$=4.05$
(from Pell\'o et al. 1999). {\bf a)} Mean spectra of the a and b
components of H5, each of them showing a strong Ly$\alpha$ emission
line at $z = 4.05$..
{\bf b)} Photometric redshift likelihood-map of H5 showing the
excellent agreement with the spectroscopic redshift, which is contained within the
region at 68\% confidence level. The shaded region on
the lower part of the Age vs. redshift map is
excluded because of age-limit considerations for the stellar population (stars cannot
be older than the age of the universe, with  $H_0$ = 50 km s$^{-1}$ Mpc$^{-1}$ and
$q_0=0$). 
{\bf c)} Restored $I_{W}$ image of H5 on the source plane at $z =
4.05$, as obtained from the lens-inversion procedure. 
}
\label{a2390}
\end{figure*}

\subsection{Photometric Redshifts and lensed sources}

Photometric Redshifts techniques in lensing clusters allow to
determine the redshift distribution and properties of high-$z$ 
and/or faint samples of lensed galaxies, sometimes invisible otherwise. 
GTs with well constrained mass distributions
enable to recover precisely the lens-corrected redshift distribution
of background sources. Thus, \zphots allow not only to select high-z
lensed sources for subsequent spectroscopic studies (as seen in
Section~\ref{TGs}), but also to obtain robust constraints on the
statistical properties of the background population of galaxies.

One of the main issues is the determination of 
deep number counts, which are strongly related to the evolution of the
spectral content of galaxies, a process driven by the star formation
history. The cumulative redshift distributions, specially in the
near-IR bands, can be used as a direct test for the scenarios of galaxy
formation (Kauffmann \& Charlot 1998; Fontana et al. 1999; 
Cole et al. 2000). Photometric Redshifts techniques 
have been recently applied to the study of galaxy properties 
up to the faintest magnitudes on the HDFs, to determine 
the star formation history at high redshift from the UV luminosity
density (Lanzetta et al. 2002), the evolution of the luminosity
functions (Bolzonella et al. 2002), and also to
analyse the stellar population and the evolutionary properties of
distant galaxies (e.g.\ SubbaRao et al.\ 1996; Gwyn \&
Hartwick 1996; Sawicki et al.\ 1997; Connolly et al.\
1997; Pascarelle et al.\ 1998; Giallongo et al.\
1998; Fern\'andez-Soto et al.\ 1999; Poli et al.\
2001), or to derive the evolution of the clustering properties
(Arnouts et al.\ 1999; Magliocchetti \& Maddox
1999, Arnouts et al.\ 2002). All these developments can be easily
applied to lensing clusters for which deep multicolor photometric
surveys are available.  

   A Monte-Carlo approach has been recently adopted by different 
authors to compute statistical quantities from photometric redshifts, 
such as the evolution of luminosity functions in deep photometric 
surveys, both in blank and strong lensing fields (SubbaRao et
al. 1996; Dye et al. 2001). 
An interesting method was proposed by Bolzonella et al. (2002) to
account for the non-gaussianity of the redshift probability functions,
and specially to include degenerate solutions in redshift. 

An interesting issue for \zphots associated to GTs for the
spectroscopic follow-up of faint amplified sources is the optimization
of the survey, i.e. selecting the best spectral domain in the visible
or near-IR bands in order to observe the strongest spectral features.
An additional benefit of \zphots is that this technique efficiently
contributes to the identification of objects with ambiguous spectral
features, such as isolated emission lines. An example of this is given
in Campusano et al. (2001).

An important concern for the different \zphot methods is that the training set 
(or the control set in the case of SED fitting) is constituted by 
the brightest objects, for which a spectroscopic 
measure of the redshift could be obtained. There is no guarantee 
that we are dealing with the same type of objects in the photometric 
(faint) and the spectroscopic (bright) samples. Thanks to the
magnification factor, cluster lenses allow the calibration of \zphots
beyond the spectroscopic limits in standard surveys, i.e. enlarging
the training/control samples towards the faintest limits in magnitude.

Another interesting and still poorly explored domain is the
combination between photometric and lensing redshifts. Indeed, lensing
inversion and \zphot techniques produce independent probability
distributions for the redshift of amplified sources. Therefore, 
the combination of both methods shall provide a robust way
to determine the redshift distribution of the faintest sources, at
least at $z \ltapprox 1.5-2$, where the lensing inversion technique
produces reliable results (Ebbels et al. 1998). In general, 
the \zphot determination is
more accurate than the $z_{lens}$ for individual objects, 
in particular at $z \gtapprox 1.5$. Nevertheless, the combination 
of both probability distributions is particularly useful when a 
degenerate solution appears in \zphot.

\subsection{Photometric Redshifts and lensing clusters}

Cluster lenses become useful GTs when their mass distribution is
highly constrained by multiple images, revealed by high resolution HST
images or multicolor photometry. The \zphot technique is particularly
useful to identify objects with similar SEDs, and
thus to check the compatibility of multiple image configurations. 
A recent example has been published by Gavazzi et al. (2003), 
who derived the redshifts of the two main radial and tangential arc
systems in the cluster MS2137-23 ($z_{cluster}=0.313$) from $UBVRIJK$ 
photometry, in order to improve the modeling 
of the central dark matter distribution. In this case, the redshift of 
the two arc systems is \zphot $= 1.6\pm 0.1$, in excellent agreement 
with the subsequent spectroscopic determination. 
\zphots 
can be used advantageously when photometric data span a large 
wavelength range, and particularly if near-IR data are available,
because this allows to obtain accurate redshifts in the sensitive 
region of $0.8 \le z \le 2$.

Photometric redshifts are particularly useful to derive the mass from weak shear
analysis. We can use \zphots to discriminate between cluster and foreground
galaxies to scale the lensing distance modulus in order
to compute the mass from the gravitational convergence.
The average convergence $\kappa \equiv \Sigma / \Sigma_{cr}$,
which corresponds to the ratio between the surface mass density and the critical
value for lensing, may be obtained as a function of the radial distance
$\theta$ using different methods (see Mellier 1999 for a review).
The mass within an aperture $\theta$ is given by

\begin{equation}
M (<\theta)  = \kappa(<\theta) \ \Sigma_{cr} \cdot \pi \left(\theta
D_{ol}\right)^2  \nonumber
  =  \kappa\left(<\theta\right) \ \theta^2 \,\frac{c^2}{4G}
\left<\frac{D_{ls}}{D_{os}D_{ol}}\right>^{-1}
\end{equation}

\noindent where $D_{ij}$ are the angular size distances between the 
cluster ($l$), the observer ($o$) and the source ($s$), and 
$\kappa(<\theta)$ is the averaged convergence
within the radius $\theta$. Using the redshift distribution N(\zphot) 
computed through a
suitable filter set, the mean value of
$\left<\frac{D_{ls}}{D_{os}D_{ol}}\right>$
can be computed, thus leading to a fair estimate of the mass. This method has
been recently applied by Athreya et al. (2002) to determine the mass
of the lensing cluster MS1008-1224. King et al. (2002) have used a
similar procedure in A1689 to select background galaxies for their 
first detection of weak gravitational shear at infrared wavelengths.
Because of the small projected surface across the redshift space, the effective
surface which is ``seen" through a cluster lens is relatively small, thus
producing a strong variance from field to field. Obtaining the $N(z_{phot})$
distribution for each cluster helps improving the mass determination.

   Another different strategy for cluster mass reconstruction using lens
magnification and \zphot techniques was proposed by Dye et al. (2001),
with an application to Abell 1689. These authors used for the first time the lens
magnification inferred from the luminosity function of background
sources when scaling the convergence to real lens mass. 

Photometric redshifts are also particularly useful to improve the 
detection of clusters in wide-field surveys, as well as to properly
identify the visible counterpart of complex and multiple lenses. 
Examples of composite
lenses recently identified by \zphot are MS1008-1224 (Athreya et
al. 2002), where a secondary lens exists at $z \sim 0.9$ in addition
to the main cluster, and the Cloverleaf field (Kneib et al. 1998).
Thompson et al. (2001) have used a deep optical and near-infrared
survey to properly identify a massive cluster of galaxies at z$=1.263$
with the X-ray source RX J105343+5735.
Photometric redshifts techniques have been recently applied to the 
ESO Distant Cluster Survey (EDisCS, White et al. 2003, in preparation),
which contains two different samples of clusters: a low-$z$ sample
at $z \sim 0.5$, for which BVIK photometry is available, and a
high-$z$ sample at $z \sim 0.8$, with VRIJK photometry. The first
comparison between spectroscopic and photometric redshifts (on $\sim$ 500
spectra) confirms the excellent behaviour of \zphot, the typical
dispersion beeing $\sigma_z \sim 0.06$ and $\sigma_z \sim 0.08$ for
the low and the high-$z$ samples respectively. These results
illustrate the typical accuracy that we can expect with well suited
photometric data, at least up to a redshift of $\sim 1$. 

\section{Looking for the first sources with Gravitational Telescopes}
\label{popIII}

\subsection{Recent developments in PopIII studies}

One of the main possible applications of GTs is the research and
study of the very first stars and galaxies forming from pristine
matter in the early Universe, the so-called Population III objects
(cf.\ review of Loeb \& Barkana 2001; Weiss et al. 2000; Umemura \&
Susa 2001).
These sources constitute the first building blocks of galaxies.
Their detection still remains one of the major challenges of present day
observational cosmology. The recent WMAP results seem to place these 
PopIII objects at redshifts up to $z \sim$ 10--30 (Kogut et al. 2003), 
a redshift
domain where the most relevant signatures are expected in the near-IR
window. During the last years, modeling efforts have been motivated by
future space facilities such as JWST,  which should be able to observe
these objects at redshifts $z$ up to 30. Nevertheless, the detection 
and first studies on the physical properties of Population III objects 
could likely be started earlier using ground-based 10m class
telescopes, and well suited near-IR facilities, such as Isaac/VLT, 
the multi-object spectrograph EMIR at GTC ($\sim$ 2006), or the future 
KMOS for the second generation of VLT instruments ($\sim$ 2008), at
least up to z$\sim$ 18.

According to modeling results, several observational signatures 
of Pop III stars and galaxies (i.e. ensembles/clusters of Pop III
stars) are expected. In summary:

\begin{itemize}

\item Strong UV emission and characteristic recombination lines of hydrogen
and He~II, especially Lyman $\alpha$ and HeII1640 (Tumlinson \& Shull 2000,
Bromm et al. 2001, Schaerer 2002, 2003).

\item Mid-IR molecular hydrogen lines at 2.12 $\mu$m and longer wavelengths
formed in cooling shells (Ciardi \& Ferrara 2001).

\item Individual supernovae whose visibility in the rest-frame optical and
near-IR could be enhanced due to time dilatation (Miralda-Escude \& Rees 1997,
Heger et al. 2001).

\item High energy neutrinos from Pop III gamma-ray bursts possibly 
associated with  fast X-ray transients (Schneider et al. 2002).

\end{itemize}

   Rest-frame UV stellar and nebular continuous and recombination line
emission represent the largest fraction of the energy emitted by
Population III objects, which are generally thought to be
predominantly massive or very massive stars, up to $\sim 1000
M_{\sun}$ (e.g. Abel et al. 1998;
Bromm et al.  2002; Nakamura \& Umemura 2001; Nakamura \& Umemura
2002). The predicted rest-frame
UV to optical spectra of Pop III galaxies including the strongest emission
lines (Lyman $\alpha$, HeII$\lambda$1640 and others) have recently
been computed by Schaerer (2002, 2003). 
The existence of a significant population of PopIII, or extremely metal-poor
objects, at redshifts $z \la$ 10 is supported by
several indications, including observations of Ly-$\alpha$ emitters with
unusually large equivalent widths in the LALA survey (Malhotra \& Rhoads 2002)
and the recent simulations of Scannapieco et al.\ (2003) accounting for
the non-homogeneous chemical evolution and pollution of the Universe.

\begin{figure}
\centerline{
\vbox{
\psfig{file=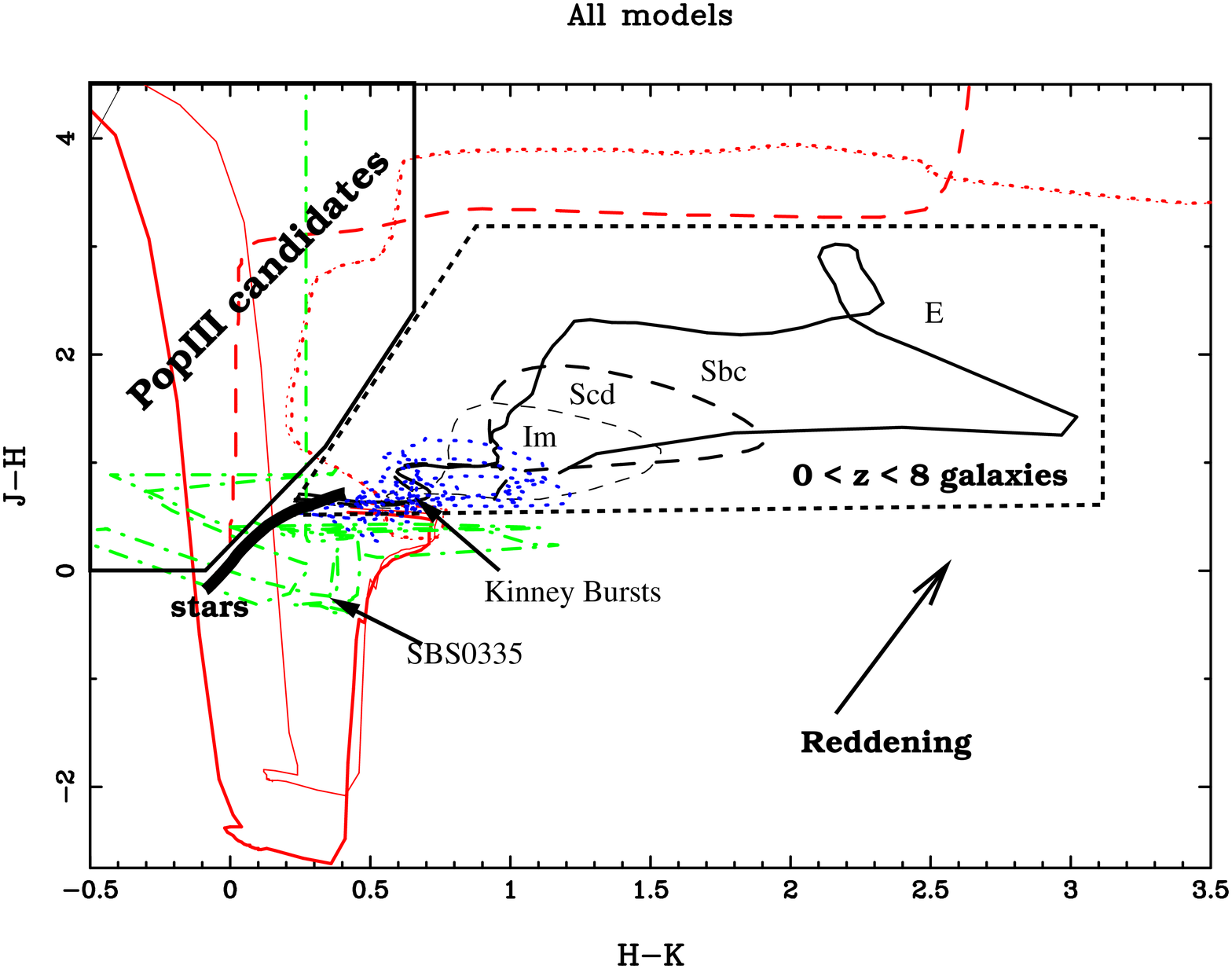,width=0.70\textwidth}
\psfig{file=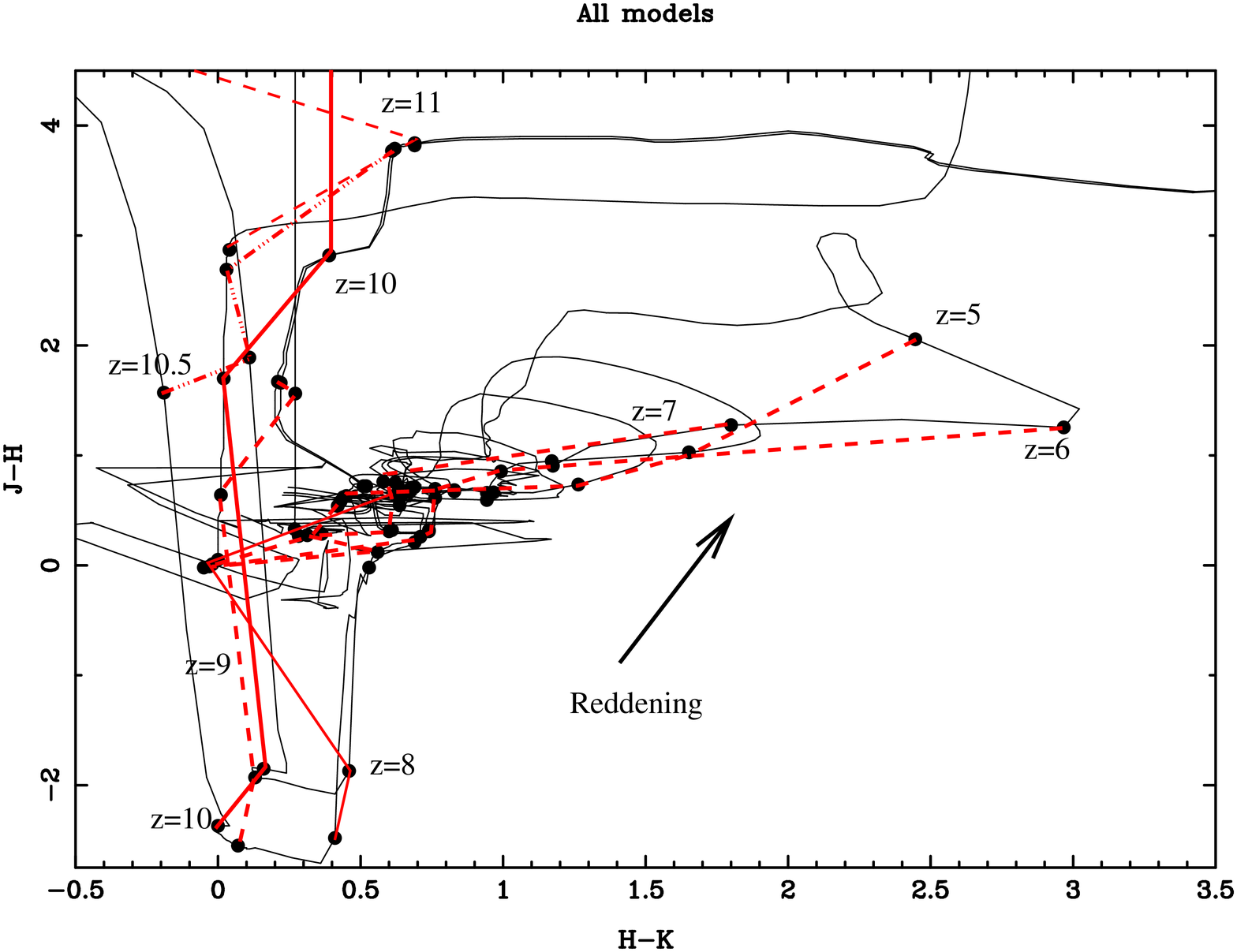,width=0.70\textwidth}}}
\caption{{\bf Top: } J-H versus H-K' color-color diagram (Vega system) for Pop III
candidates. Various models for Pop III objects are presented
over the interval $z \sim$ 5 to 12, with
different fractions of Lyman $\alpha$ emission entering the
integration aperture: 100\% (thick solid lines), 50\% (thin solid
lines), and 0\% (thin dotted lines). Thick dotted lines correspond to 
colors obtained within an integration aperture limited to the
non-resolved core of the source, and Lyman~$\alpha$ emission coming
from an extended halo (Loeb \& Rybicki 1999).
Positions of stars and normal galaxies up to z $\le$ 8
are also shown for comparison, including starbursts templates 
(SB1 and SB2, from Kinney et al. 1993 -blue dotted lines-,
and the low metallicity galaxy SBS0335-052 -green dash-dotted
lines-). The shift direction induced by reddening is also indicated.
See text for more details.
{\bf Bottom: } Same color-color diagram as in Top Figure, showing the
location of the iso-z lines: z $<$ 8 (thick dashed lines), z=8 (thin solid
line), z=9 (thick dashed line),
z=10 (thick solid line), z=10.5 (thick dotted-dashed line) and z=11
(thin dashed line).
}
\label{popIII_cc}
\end{figure}

\begin{figure}
\centerline{\psfig{file=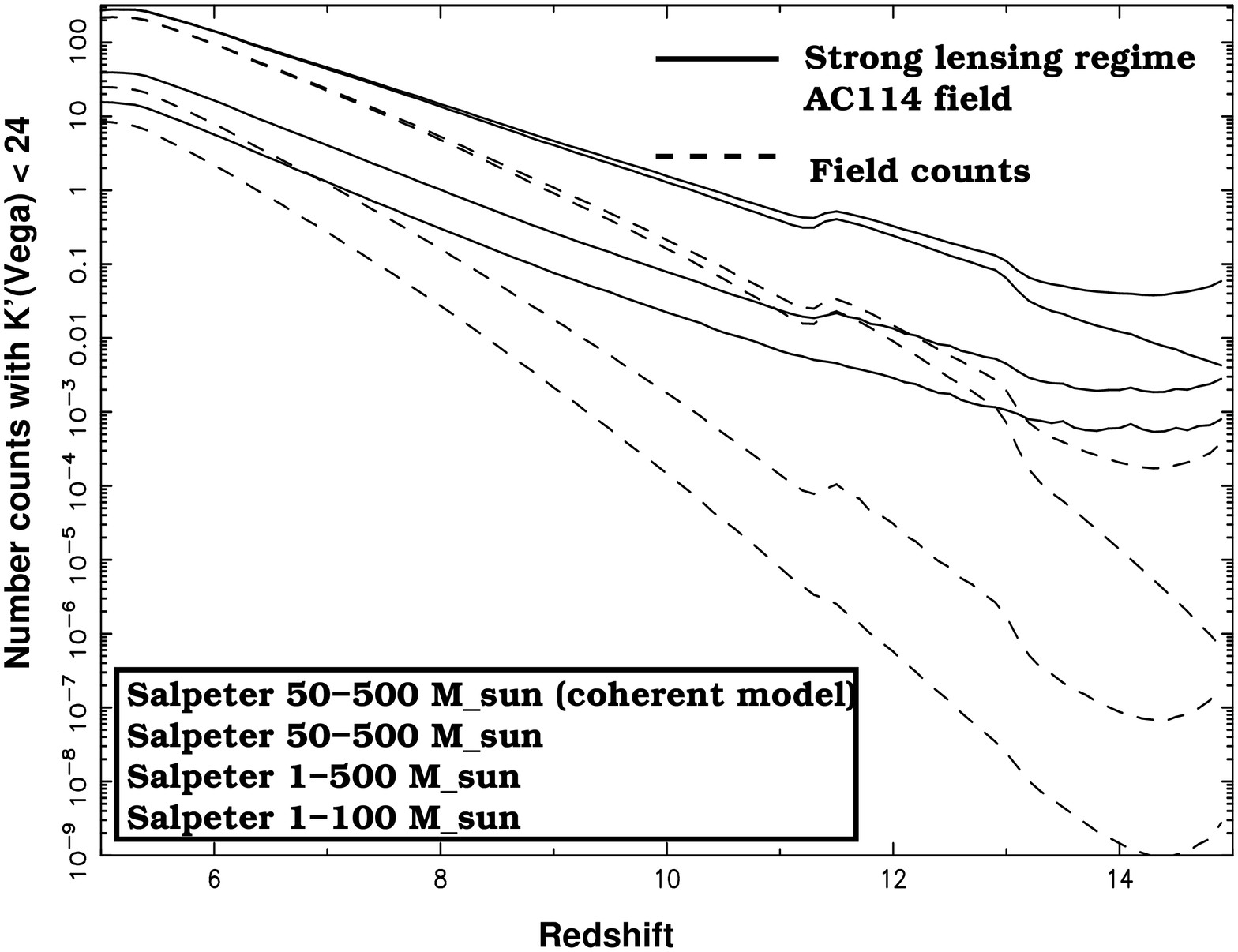,width=1.00\textwidth,angle=270}}
\caption{Expected number counts of PopIII objects with K'(Vega) $<$ 24,
in the field of view of Isaac, per 0.1 redshift bin interval, using
different assumptions for the IMF (from Richard et al., in
preparation). The typical results expected in
the strong lensing regime (solid lines), towards the
lensing cluster AC114, are compared to the
corresponding ones in blank fields surveys (dashed lines),
for the same photometric depth. The gain in the z $\sim 8-10 $ domain
when taking advantage of the lensing amplification is typically a
factor of 10 (see text for details).}
\label{popIII_counts}
\end{figure}

\subsection{PopIII observational signatures}

Simulations have been carried out using the recent models by Schaerer
(2002) to demonstrate the feasability of this
scientific case in view of the future near-IR facilities mentioned
above. The aim is to determine a well suited identification criteria
for PopIII galaxies, and to study the feasability of spectroscopic 
observations in the near-IR domain. Some preliminary results have 
been shown in Schaerer \&
Pell\'o (2000) and Pell\'o \& Schaerer (2002).

The SED of PopIII sources is dominated by the
nebular continuous emission  at $\lambda \ge 1400$ \AA, and several
strong emission lines are present, such as Lyman $\alpha$, 
HeII$\lambda$1640, HeII$\lambda$3203, HeII$\lambda$4686 and others.
We have considered different IMFs and ages for the stellar population,
as well as two different star formation regimes (starburst and continuous
star formation), following the prescriptions given by Schaerer (2002).
Lyman series troughs (Haiman \& Loeb 1999), and Lyman forest following the
prescription of Madau (1995) are included. The reionization
redshift is assumed to be $\sim$6. The virial radius is of the 
order of a few kpc for these sources, and thus we
consider that sources are unresolved on a 0.3'' scale, with
spherical symetry. Different series of simulations have been 
performed to account for different hypothesis
for the extended Ly~$\alpha$ halo (cf. Loeb \& Rybicki 1999).

   Figure~\ref{popIII_cc} displays the J-H versus H-Ks color-color 
diagram (Vega system) for Pop III objects obtained from these
simulations, showing 
the expected position of objects at different redshifts over the
interval of $z \sim$ 5 to 12, and the 
location for the most promising Pop III candidates.
Various models for Pop III objects are presented in
Figure~\ref{popIII_cc}. 
In particular, models are computed assuming different fractions 
of Lyman $\alpha$ emission entering the
integration aperture: 100\% (thick solid lines), 50\% (thin solid
lines), and 0\% (thin dotted lines). The expected colors obtained within
an integration aperture corresponding to the core (non-resolved)
source, whith Lyman~$\alpha$ emission coming from an extended halo
(Loeb \& Rybicki 1999), are displayed by thick dotted lines.
The colors expected for a pure
stellar population are given for comparison as dashed thick lines.
We used a Salpeter IMF to derive the synthetic spectra in all cases.
The location of stars and normal galaxies at $0 \le z \le 8$ is also
given for comparison, as well as 
starbursts templates (SB1 and SB2, from Kinney et al. 1993,
and the low metallicity galaxy SBS0335-052). The strong UV continuum
and the presence of prominent emission lines allow to discriminate
between PopIII objects with $z \gtapprox 8$ and normal galaxies or
stars. Even ``normal'' galaxies at $z \gtapprox 8$ will 
exhibit redder H-K colors (which trace the restframe UV slope) 
as compared to PopIII.
Because of the limiting
magnitudes in the different filters, most PopIII candidates are
expected to fall within the ``PopIII candidates'' region.
This diagram illustrates that
PopIII objects with strong ionising continua can well be separated
from "normal" objects provided that ultra-deep JHK photometry with 
sufficient accurary is available.

We have also computed the expected S/N ratios for the
main emission lines, as seen through a 10m telescope, using a medium
resolution spectrograph (R$\sim 3000-5000$). 
According to our results, Lyman $\alpha$ can easily be detected with a
good S/N over the redshift
intervals z $\sim$ 8 to 18, with some gaps, depending on the spectral
resolution (OH subtraction) and atmospheric transmission. A joint
detection with HeII $\lambda$1640, the strongest HeII line, is possible
for z $\sim$  5.5-7.5 (Lyman $\alpha$ in optical domain), z $\sim$
8-14 with both lines
in the near-IR, again with some gaps. The typical line fluxes for the
HeII $\lambda$1640 line range between $10^{-17}$ and a few $10^{-18}$
erg/s/cm2, for a fiducial stellar mass halo of $10^7$ M$_{\sun}$,
corresponding to a collapsing dark matter halo of $2 \times 10^8 M_{\sun}$.
The detection of both HeII $\lambda$1640 and Ly~$\alpha$ allows
one e.g. to obtain a measure of the hardness of the ionising flux
which constrains the upper end of the IMF and the age of Pop III
systems.

 Thus, according to our results, efficient photometric
and spectroscopic observations of these features should be possible in the
near-IR domain, thanks to the future imaging and spectroscopic facilities,
for a large number of sources, thus allowing to derive statistically
significant conclusions about their formation epoch and physical
properties. Broad-band photometry is quite insensitive to changes in
the IMF, and broad-band colors do not allow
to constrain physical properties such as the IMF (i.e. the mass
range of Pop III stars), ages, etc., but could be useful to identify
the sources on ultra-deep photometric surveys. Spectroscopy
is needed to study the physics of these objects.

\subsection{A search for PopIII candidates behind lensing clusters}

PopIII objects could be detected from deep near-IR
photometry based  on a measurement of two colors with accuracies of
the order of ~0.3 mag. Ideal fields for the first studies are lensing
clusters of galaxies with areas of strong gravitational amplification.
A devoted and prototype observational program is presently going on,
aimed at the detection of Population III objects at
high $z$ using GTs. 

During the last year we were granted ISAAC/VLT time 
to obtain ultra-deep JHK' images of a gravitational lensing clusters 
in order to search for PopIII candidates
using our above mentioned color-color criteria. Two lensing clusters
were observed up to now: AC114 ($z=0.31$), images obtained late August
2002, and Abell 1835 ($z=0.253$), observed in March 2003. 
At present, only data on AC114 have been analyzed. Images were
obtained under excellent seeing (0.4-0.6") and photometric conditions. The
limiting magnitudes attained (3$\sigma$ / 4pixels) are J$=25.7$, H$=24.7$ and
K'$=24.1$ in the Vega system. From the first 
investigations on AC114, several ($\sim 10$) candidate PopIII galaxies
are found in the selection window, with magnitudes $K' \sim$ 23--24 (Vega system) and
estimated redshifts of $z \sim$ 8--11. All these objects are
R-dropouts, located close to the theoretical position with respect to
the critical lines. A few of them are brighter than $H \sim$ 24
(including 1--2 mag amplification). They constitute the first
spectroscopic targets for spectroscopic follow-up with ISAAC/VLT in
the subsequent months.

   The expected number of Pop III objects and primordial QSOs in
blank fields has been
derived by several authors. E.g. in a comprehensive study Ciardi et
al. (2000) show that at z $>$ 8 naked stellar clusters, i.e. objects which
have completely blown out their ISM, and thus avoid local chemical
enrichment, dominate the population of luminous objects.
Although pilot studies have recently started to explore
the possible formation  of dust in Pop III objects (Todini \& Ferrara
2001), the effect is generally neglected. Based on such assumptions, Oh
et al. (2001) have calculated the predicted number of Pop III objects
detectable in HeII lines with JWST for a one day integration time.
Their estimate yields between ~ 60 and 4500
sources in a 10'x10' field of view, depending on
the model parameters. These estimates are still highly uncertain and
model dependent.

In Figure~\ref{popIII_counts} we compare the expected 
number counts of PopIII objects with K'$<$ 24 in our GT ongoing survey,
within the field of view of Isaac, per 0.1 redshift bin interval, using
different assumptions for the IMF. A simple Press-Schechter
formalism for the abundance of halos
and standard $\Lambda$CDM cosmology were considered in these 
simulations, as well as a conservative fixed fraction of the baryonic 
mass in halos converted into stars, assuming that all stars are
PopIII, without chemical evolution (Richard et al., in preparation). 
Thus, number counts shown in
Figure~\ref{popIII_counts} have to be considered as a first order
estimate, the important point beeing the comparison between 
lensing and blank fields.
The typical results expected in the strong lensing regime, towards the
lensing cluster AC114, are compared to the corresponding ones in blank 
fields surveys,
for the same photometric depth. The gain in the z $\sim 8-10 $ domain
when taking advantage of the lensing amplification is typically a
factor of 10. 
The expected density of primordial quasars could
be similar to that of PopIII galaxies (Oh et al. 2001). Thus,
multiplexing is needed in blank and lensing fields to allow highly 
efficient observations of relevant samples of Pop III objects. Also, 
because the observational signatures of primordial quasars are
expected to be similar to those of genuine Pop III stars, 
a relatively high resolution is needed to obtain line profiles.

\section{Conclusions}
\label{conclusions}

Gravitational Telescopes still remain a unique tool in deep universe
studies. The recent developments in the sub-millimetric and
mid-infrared domains, the search for extremely faint EROs and
primordial galaxies, and the detailed study of the spectrophotometric
and morphological properties of galaxies beyond the spectroscopic
limits argue in favour of this technique.  

Recent results obtained by different authors on the physical
properties of faint lensed galaxies (such as AC114-S2, AC114-A2, or
the $z \sim 3$  sources in 1E0657-558) 
suggest that high-$z$ objects of different luminosities
were subjected to different histories of star formation. But these 
results are based on a dramatically small number of objects. 
Enlarged samples of galaxies in both redshift and luminosity,
specially in the redshift range 1.5 $\ltapprox$ z $\ltapprox$ 6,
are urgently needed to set strong constrains on galaxy evolution
scenarios. The systematic use of GTs to access the most distant and
faintest population of galaxies can help achieving this goal. 

   An important issue for future spectroscopic studies of very high-z 
primordial galaxies is the strategy for target selection. Photometric
redshifts and various color-selection techniques have demonstrated
their ability to produce samples of high-z objects. Even young Pop III 
sources or extremely metal-deficient starbursts are expected to show distinct 
characteristics in their near-IR colors compared to
``normal'' galaxies at any redshift. GTs with areas of strong
gravitational amplification are the ideal fields for the first
prospective studies. Thanks to its multiplexing, 
spectral resolution and wide field-of-view capabilities, the
forthcoming next generation of ground-based near-IR spectrographs, 
such as EMIR at {\it Grantecan}
({\tt http://www.iac.es/gtc/index.html})
or the future KMOS for the VLT, are the suited instruments to
complement JWST performances and to
start exploring the formation epoch of the first stars in galaxies.

\acknowledgements
Part of this work was supported by the
French {\it Centre National de la Recherche Scientifique}, by the
TMR Lensnet ERBFMRXCT97-0172 (http://www.ast.cam.ac.uk/IoA/lensnet), 
by the French {\it Conseil R\'egional de la Martinique}, and
the ECOS SUD Program.

\label{page:last}


\begin{references}


\reference Abel, T., Anninos, P.A., Norman, M.L., Zhang, Y., 1998, 
\apj{508}{518}

\reference Adelberger, 
K.~L.~\& Steidel, C.~C.\ 2000, 
\apj{544}{218} 

\reference Altieri, B., Metcalfe, L., Kneib, J. P.,  ~et al.\ 
1999, 
\aap{343}{L65} 

\reference Arnouts, S., Cristiani, S., Moscardini, L.,
Matarrese, S., Lucchin, F., Fontana, A. \& Giallongo, E., 1999,
\mnras{310}{540}

\reference Arnouts, S., Moscardini, L., Vanzella, E.,
Colombi, S., Cristiani, S., Fontana, A., Giallongo, E., Matarrese,
S. \& Saracco, P., 2002, 
\mnras{329}{355}

\reference  Athreya, R.~M.,
Mellier, Y., van Waerbeke, L., Pell\'o, R., Fort, B., \& Dantel-Fort, M.\ 2002, 
\aap{384}{743}

\reference Baum, W.A. 1962, 
IAU Symposium n. 15, Macmillan Press, New
York, p.390

\reference Ben\'{\i}tez, N. 2000, 
\apj{536}{571}

\reference B{\' e}zecourt, J., Soucail, G., Ellis, 
R.~S., \& Kneib, J.-P.\ 1999, 
\aap{351}{433} 

\reference Blain, 
A.~W., Smail, I., Ivison, R.~J., \& Kneib, J.-P.\ 1999, 
\mnras{302}{632} 

\reference Blain, 
A.~W., Kneib, J.-P., Ivison, R.~J., \& Smail, I.\ 1999, 
\apj{512}{L87} 

\reference Bolzonella M., Miralles J.M., Pell\'o R., 2000,
\aap{363}{476}

\reference Bolzonella M., Pell\'o R., Maccagni D., 2002, 
\aap{395}{443} 

\reference Bromm, V. , Kudritzki, R.P., Loeb, A., 2001, 
\apj{552}{464}

\reference Bromm, V., Coppi, P.S., Larson, R.B., 2002, 
\apj{564}{23}

\reference  Campusano, L.~E.,
Pell\'o, R., Kneib, J.-P., Le Borgne, J.-F., Fort, B., Ellis, R.,
Mellier, Y., \& Smail, I., 2001, 
\aap{37}{394}

\reference Casertano, S., de Mello, D., Dickinson, M., et
al., 2000, 
\aj{120}{2747}

\reference Chapman, S. C., Scott, D., Steidel, C. C. et al. 2000, 
\mnras{319}{318} 

\reference Ciardi, B., Ferrara, A., Governato, F., Jenkins, A., 2000,
\mnras{314}{611} 

\reference Ciardi, B., Ferrara, A., 2001, 
\mnras{324}{648}

\reference Cole, S., Lacey, C., Baugh, C. \& Frenk, C. 2000,
\mnras{319}{168}

\reference Connolly, A.J., Csabai, I., Szalay, A.S., Koo, D.C., Kron, R.G.,
Munn, J.A., 1995, 
\aj{110}{2655}

\reference Connolly, A. J., Szalay, A. S., Dickinson, M.,
SubbaRao, M. U., Brunner, R. J., 1997, 
\apj{486}{L11}

\reference Contini, T., Treyer, M.~A., Sullivan, M., \& Ellis, R.~S.\
2002, 
\mnras{330}{75} 

\reference Couch, W.J., Ellis, R.S., Godwin, J., Carter, D., 1983,
\mnras{205}{1287}


\reference  Dickinson, M., Papovich, C., Ferguson, H. C., 2001,
Proceedings of the ESO Symposium, ``Deep Fields'',
ed. S. Cristiani (Berlin: Springer), astro-ph/0105086.

\reference  Dye, S., Taylor, A. N., Thommes, E. M.,
Meisenheimer, K., Wolf, C. \& Peacock, J. A., 2001, 
\mnras{321}{685}

\reference Ebbels, T. M. D.,  Le Borgne, J. -F.,
Pell\'o, R., Ellis, R. S., Kneib J. -P., Smail, I.; Sanahuja, B., 1996,
\mnras{281}{L75}

\reference Ebbels, T., Ellis, R.,
Kneib, J., Le Borgne, J., Pell\'o, R., Smail, I., \& Sanahuja, B., 1998,
\mnras{295}{75}

\reference Ellis, R.~S., Colless, 
M., Broadhurst, T., Heyl, J., \& Glazebrook, K.\ 1996, 
\mnras{280}{235} 

\reference Ellis, R., Santos, M.~R., Kneib, J., \& Kuijken, K.\ 2001, 
\apj{560}{L119} 

\reference Erb, D. K., Shapley, A. E., Steidel, C. C., Pettini, M.,
Adelberger, K. L., Hunt, M. P., Moorwood, A. F. M., Cuby, J.-G., 2003,
ApJ in press, astro-ph/0303392

\reference Fern\'andez-Soto, A., Lanzetta, K.M. \& Yahil,
A. 1999, 
\apj{513}{34}

\reference Fort B. \& Mellier Y., 1994, 
A\&AR 5, 239

\reference Fontana A., Menci N., D'Odorico S., Giallongo E.,et
al. 2001, 
\mnras{310}{L27}

\reference Franx, M., Illingworth, 
G.~D., Kelson, D.~D., van Dokkum, P.~G., \& Tran, K.\ 1997, 
\apj{486}{L75} 

\reference Frye, B.~\& Broadhurst, T.\ 1998, 
\apj{499}{L115} 

\reference Furusawa, H., Shimasaku, K., Doi, M., Okamura, S., 2000,
\apj{534}{624}

\reference Gavazzi, R., Fort, B., Mellier, Y., Pello, R., Dantel-Fort,
M., 2003,
A \& A in press, astro-ph/0212214.

\reference Giallongo, E., D'Odorico,, S., Fontana, A.,
Cristiani, S., Egami, E., Hu, E. \& McMahon, R. G., 1998, 
\aj{115}{2169}

\reference Gwyn, S. D. J. \& Hartwick F. D. A., 1996, 
\apj{468}{L77}

\reference Haiman, Z., Loeb, A., 1998, 
\apj{503}{505}

\reference Hammer, F., Gruel, N., 
Thuan, T.~X., Flores, H., \& Infante, L.\ 2001, 
\apj{550}{570} 

\reference Heger, A., Woosley, S. E., Baraffe, I., Abel, T., 2001,
Proc. MPA/ESO/MPE/USM Joint Astronomy Conference "Lighthouses of the
Universe: The Most Luminous Celestial Objects and their use for
Cosmology", p. 369

\reference  Hu, E.~M., Cowie, L.~L., 
McMahon, R.~G., Capak, P., Iwamuro, F., Kneib, J.-P., Maihara, T., \&
Motohara, K.\ 2002, 
\apj{568}{L75} 

\reference Ivison, R.~J., Smail, 
I., Barger, A.~J., Kneib, J.-P., Blain, A.~W., Owen, F.~N., Kerr, T.~H., \& 
Cowie, L.~L.\ 2000, 
\mnras{315}{209} 

\reference King, L.~J., Clowe, D.~I., 
Lidman, C., Schneider, P., Erben, T., Kneib, J.-P., \& Meylan, G.\
2002, 
\aap{385}{L5} 

\reference Kneib, J.-P.,
Mathez, G., Fort, B., Mellier, Y., Soucail, G. \& Longaretti, P.-Y.,
1994, \aap{286}{701}

\reference  Kneib, J.-P.,
Ellis, R.S., Smail, I., Couch, W.J., Sharples, R., 1996,
\apj{471}{643}

\reference  Kneib, J.-P.,
Alloin, D., \& Pell\'o, R.\, 1998, 
\aap{339}{L65}

\reference Kauffmann, G. \& Charlot, S. 1998, 
\mnras{297}{L23}

\reference Kobulnicky, 
H.~A.~\& Zaritsky, D.\ 1999, 
\apj{511}{118} 

\reference  Kogut, A., Spergel, D. N.,  Barnes, C.,
Bennett, C. L., Halpern, M., G. Hinshaw , G. , Jarosik ,N. 
Limon , M., Meyer, S. S.,  Page , L., Tucker, G., 
Wollack, E.,  Wright, E. L., 2003, 
submitted to ApJ, astro-ph/0302213.

\reference Koo, D.C., 1985, 
\aj{90}{418}

\reference Labb{\' e}, I.~et al.\ 2003, 
\aj{125}{1107}

\reference Lanzetta, K.M., Yahil, A., Fern\'andez-Soto, A. 1996,
Nature 381, 759

\reference Lanzetta, K.~M., 
Yahata, N., Pascarelle, S., Chen, H., \& Fern{\' a}ndez-Soto, A.\
2002, 
\apj{570}{492} 

\reference Le Borgne, D.~\& Rocca-Volmerange, B., 2002,
\aap{386}{446}

\reference Leitherer, C. 1999, Chemical Evolution from Zero to High Redshift,
Proceedings of ESO Workshop, Garching, Germany, 14-16 October
1998. Eds. R. Walsh, M. R. Rosa. Berlin: Springer-Verlag , 204

\reference Lemoine-Busserolle, M., Contini, T., Pell{\' o}, R., Le Borgne, J.-F., 
Kneib, J.-P., \& Lidman, C.\ 2003, 
\aap{397}{839} 

\reference Lilly, S.~J., Le Fevre, 
O., Crampton, D., Hammer, F., \& Tresse, L.\ 1995, 
\apj{455}{50} 

\reference Loeb, A. \& Barkana, R., 2001, 
\araa{39}{19}

\reference Madau, P. 1995, 
\apj{441}{18}

\reference Magliocchetti, M. \& Maddox, S. J., 1999, 
\mnras{306}{988}

\reference Malhotra, S.~\& Rhoads, J.~E.\ 2002,  
\apj{565}{L71}

\reference Mehlert, D., Seitz, S., Saglia, R. P., Appenzeller, I.,
Bender, R., Fricke, K. J., Hoffmann, T. L., Hopp, U., Kudritzki,
R.-P., Pauldrach, A. W. A., 2001,
\apj{379}{96}

\reference Mehlert, D.~et al.\ , 2002, 
\aap{393}{809} 

\reference Melbourne, J.~\& Salzer, J.~J.\ 2002, 
\aj{123}{2302} 

\reference Mellier, Y., Fort, B., 
Soucail, G., Mathez, G., \& Cailloux, M.\ 1991, 
\apj{380}{334} 

\reference Mellier, Y., 1999, 
ARA \& A 37, 127 

\reference Miralda-Escude, J., Rees, M.J., 1997, 
\apj{478}{L57}

\reference Mobasher, B., Rowan-Robinson, M., Georgakakis, A., Eaton, N.,
1996, 
\mnras{282}{L7}

\reference Mouhcine, M.~\& Contini, T.\ 2002, 
\aap{389}{106} 

\reference Nakamura, F., Umemura, M., 2001, 
\apj{548}{19}

\reference Nakamura, F., Umemura, M., 2002,
\apj{569}{549}

\reference Natarajan, P., Kneib, J., Smail, I., \& Ellis, R.~S.\ 1998, 
\apj{499}{600} 

\reference Oey, M.~S.~\& Kennicutt, R.~C.\ 1993, 
\apj{411}{137} 

\reference Oh, S.~P., Haiman, 
Z., \& Rees, M.~J.\ 2001, 
\apj{553}{73} 

\reference Pascarelle, S. M., Lanzetta, K. M. \&
Fern\'andez-Soto, A., 1998, 
\apj{508}{L1}

\reference Pell{\' o}, R., Kneib, J.-P., Le Borgne J.-F., Bezecourt,
J., Ebbels, T. M., Tijera, I., Bruzual, G., Miralles, J. M., Smail,
I., Soucail, G., Bridges, T. J., 1999, 
\aap{346}{359}

\reference Pell\'o, R., D. Schaerer, 2002,
Proceedings of  "Science with the GTC", Granada, Spain, 
(astro-ph/0203203), to be published in Revista Mexicana de Astronomia y
Astrofisica.

\reference Pettini, M., Kellogg, M., Steidel, C.~C., Dickinson, M.,
Adelberger, K.~L., Giavalisco, M., 1998, 
\apj{508}{539}

\reference Pettini, M., Steidel, C.~C., Adelberger, K.~L., Dickinson, M.,
Giavalisco, M., 2000, 
\apj{528}{96}
 
\reference Pettini, M., Shapley, A.~E., Steidel, C.~C., et al., 2001, 
\apj{554}{981}

\reference Pettini, M., Rix, S.~A., Steidel, C.~C., Adelberger, K.~L.,
Hunt, M.~P., Shapley, A.~E., 2002a,  
\apj{569}{742}

\reference Pettini, M., Ellison, S.~L., Bergeron, J., Petitjean, P.,
2002b,  
\aap{391}{21}

\reference Poli, F., Menci, N., Giallongo, E., Fontana, A.,
  Cristiani, S., D'Odorico, S., 2001 , 
\apj{551}{L45}

\reference Richer, M.~G.~\& McCall, M.~L.\ 1995, 
\apj{445}{642} 

\reference Rodighiero, G., Franceschini, A. \& Fasano,
G., 2001, 
\mnras{324}{491}

\reference Rudnick, G., Franx, M., Rix, H.-W., Moorwood,
A., Kuijken, K., van Starkenburg, L., van der Werf, P., R\"ottgering,
H., van Dokkum, P. \& Labbe, I., 2001, 
\aj{122}{2205}

\reference Sawicki, M. J., Lin, H. \& Yee, H. K. C., 1997,
\aj{113}{1}

\reference Scannapieco, E., Schneider, R., Ferrara, A., 2003,
Apj in press, astro-ph/0301628

\reference Schaerer, D. 2002, 
\aap{382}{28} 

\reference Schaerer, D. 2003, 
\aap{397}{527}

\reference Schaerer, D. \& Pell\'o R., 2001, 
in "Scientific Drivers for ESO Future VLT/VLTI Instrumentation",
J. Bergeron and G. Monnet, Eds.,
Springer Verlag, p.48, astro-ph/0107274 

\reference Schneider, R., Guetta, D., \& Ferrara, A.\ 2002, 
\mnras{334}{173} 

\reference Seitz, S., Saglia, R.~P., 
Bender, R., Hopp, U., Belloni, P., \& Ziegler, B.\ 1998, 
\mnras{298}{945} 

\reference Smail, I., Ellis, R.~S., 
Aragon-Salamanca, A., Soucail, G., Mellier, Y., \& Giraud, E.\ 1993, 
\mnras{263}{628}

\reference Smail, I., Couch, W.~J., Ellis, R.~S., \& Sharples, R.~M.\ 1995, 
\apj{440}{501} 

\reference Smail, I., Dressler, A., 
Kneib, J., Ellis, R.~S., Couch, W.~J., Sharples, R.~M., \& Oemler, A.~J.\ 
1996, 
\apj{469}{508} 

\reference Smail, I., 
Ivison, R.~J., \& Blain, A.~W.\ 1997, 
\apj{490}{L5} 

\reference Smail, 
I., Ivison, R.~J., Blain, A.~W., \& Kneib, J.-P.\ 1998, 
\apj{507}{L21} 

\reference Smail, 
I., Ivison, R.~J., Blain, A.~W., \& Kneib, J.-P.\ 2002, 
\mnras{331}{495} 

\reference Smith, G.~P., Smail, I., Kneib, J.-P., Czoske, O., Ebeling, H., Edge, A. C., Pell\'o, R., Ivison, R. J., Packham, C., Le Borgne, J.-F., 2002, 
\mnras{330}{1} 

\reference Soifer, B.~T., 
Neugebauer, G., Franx, M., Matthews, K., \& Illingworth, G.~D.\ 1998,
\apj{501}{L171} 

\reference Soucail, G., Mellier, 
Y., Fort, B., Mathez, G., \& Cailloux, M.\ 1988, 
\aap{191}{L19} 

\reference Soucail, G., Kneib, 
J.~P., B{\' e}zecourt, J., Metcalfe, L., Altieri, B., \& Le Borgne, J.~F.\ 
1999, 
\aap{343}{L70} 

\reference Steidel, 
C.~C., Pettini, M., \& Hamilton, D.\ 1995, 
\aj{110}{2519} 

\reference Steidel, C.~C., Giavalisco, M., 
Dickinson, M., \& Adelberger, K.~L.\ 1996, 
\aj{112}{352} 

\reference Steidel, C.~C., 
Adelberger, K.~L., Giavalisco, M., Dickinson, M., \& Pettini, M.\ 1999, 
\apj{519}{1} 

\reference SubbaRao, M. U., Connolly, A. J., Szalay,
A. S. \& Koo, D. C.,  1996, 
\aj{112}{929}

\reference Telles, E.~\& Terlevich, R.\ 1997, 
\mnras{286}{183} 

\reference Teplitz, H. I., McLean, I. S., Becklin, E. E., et al., 2000, 
\apj{533}{L65} 

\reference Thompson, D.~et al.\ 
2001, 
\aap{377}{778} 

\reference Todini, P., Ferrara, A., 2001, 
\mnras{325}{726}

\reference Trager, S.~C., Faber, S.~M., Dressler, A., \& Oemler,
A.~J.\ 1997, 
\apj{485}{92}

\reference Tumlinson, J., Shull, J.M., 2000, 
\apj{528}{L65}

\reference Wang, Y., Bahcall, N., Turner, E.L., 1998, 
\aj{116}{2081}

\reference Weiss, A., Abel, T., Hill, V., Eds., 2000, 
``The First Stars'',
MPA/ESO Workshop 1999, Garching, Spring Verlag, Heidelberg.

\reference  Williams, R. E., Blacker, B., Dickinson, M.,
et al., 1996, 
\aj{112}{1335}

\reference Umemura, M., Susa, H. Eds., 2001,''The physics of galaxy formation''
ASP Conf. Series, Vol. 222 

\reference Yee, H.~K.~C., Ellingson, 
E., Bechtold, J., Carlberg, R.~G., \& Cuillandre, J.-C.\ 1996, 
\aj{111}{1783} 

\reference Zaritsky, D., Kennicutt, R.~C., \& Huchra, J.~P.\ 1994, 
\apj{420}{87}



\end{references}
\end{document}